\documentclass[10pt, conference, compsocconf]{IEEEtran}
\IEEEoverridecommandlockouts
\ifCLASSINFOpdf
\else
\fi
\usepackage{amsmath}
\usepackage{amsfonts}
\usepackage{amsthm}
\usepackage{stmaryrd}
\usepackage{comment}
\usepackage{graphicx}
\usepackage[pdf]{pstricks}
\usepackage{xspace}
\usepackage{tikz}
\usepackage{gnuplot-lua-tikz}
\usepackage{pgfplotstable} 
\pgfplotsset{compat=1.12}
\usepackage{booktabs}

\newtheorem{mytheorem}{Theorem}

\theoremstyle{definition}

\newtheorem{mydefinition}[mytheorem]{Definition}

\newcommand{\R}{{\mathbb{R}}}
\newcommand{\Rnn}{\R_{\ge 0}}

\newcommand{\libmonaa}{${\mathtt{libmonaa}}$\xspace}
\newcommand{\monaa}{${\mathtt{MONAA}}$\xspace}

\makeatletter
\newcommand{\figcaption}[1]{\def\@captype{figure}\caption{#1}}
\newcommand{\tblcaption}[1]{\def\@captype{table}\caption{#1}}
\makeatother

\usepackage{tikz}
\usepackage{textcomp}
\usepackage{hyperref}
\usepackage{lipsum}

\newcommand\copyrighttext{%
  \footnotesize \textcopyright 2018 IEEE. 
  DOI: \href{https://doi.org/10.1109/MT-CPS.2018.00014}{10.1109/MT-CPS.2018.00014}}
\newcommand\copyrightnotice{%
\begin{tikzpicture}[remember picture,overlay]
\node[anchor=south,yshift=10pt] at (current page.south) {\fbox{\parbox{\dimexpr\textwidth-\fboxsep-\fboxrule\relax}{\copyrighttext}}};
\end{tikzpicture}%
}

\begin{document}
%
\title{MONAA: a Tool for Timed Pattern Matching with Automata-Based Acceleration}

\author{\IEEEauthorblockN{Masaki Waga\IEEEauthorrefmark{1},
Ichiro Hasuo\IEEEauthorrefmark{2}, and
Kohei Suenaga\IEEEauthorrefmark{3},}
\IEEEauthorblockA{\IEEEauthorrefmark{1}The University of Tokyo, Tokyo
Japan Email: mwaga@is.s.u-tokyo.ac.jp}
\IEEEauthorblockA{\IEEEauthorrefmark{2}National Institute of
Informatics, Tokyo, Japan Email: i.hasuo@acm.org}
\IEEEauthorblockA{\IEEEauthorrefmark{3}Kyoto University Kyoto, Japan Email: ksuenaga@fos.kuis.kyoto-u.ac.jp}}



\maketitle
\copyrightnotice


%
\IEEEpeerreviewmaketitle

\section{Monitoring}

\emph{Monitoring} over a real-time specification is an actively studied
topic with a lot of industrial applications, such as monitoring of 
simulation traces of a Simulink model, and a HILS
(\emph{hardware-in-the-loop simulation}) with a system prototype. Given a
log (a \emph{timed word} or a \emph{signal}) and a specification (a
\emph{timed automaton} (TA){}~\cite{Alur1994}, a \emph{timed regular
expression} (TRE)~\cite{Asarin2002}, or a formula in
\emph{metric temporal logic}~\cite{DBLP:journals/rts/Koymans90}), a monitor finds all the segments of the log
that satisfy the given specification. 

A monitoring procedure has the \emph{online}
property if it starts the procedure before the entire log is given. This
property is essential in monitoring a system that is currently running. 
The \emph{efficiency} of a monitoring procedure is also important because
 recent trends such as autonomous driving have drastically increased the size of logs
and the number of properties to monitor. Also, when the procedure is executed
on a cloud server, an efficient monitor can reduce the pay-as-you-go cost.

One possible formalization of such monitoring problems is given by
\emph{timed pattern matching}, and both offline and online algorithms are proposed. See~\cite{Ulus2014} and~\cite{Ulus2016}; their theoretical results have led to
their tool Montre~\cite{DBLP:conf/cav/Ulus17}. Besides this series of works, the current authors
have investigated efficient algorithms for timed pattern matching with
automata-based acceleration: see~\cite{DBLP:conf/formats/WagaAH16} and~\cite{DBLP:conf/formats/WagaHS17}. The acceleration in our algorithms is based on the idea of \emph{skipping} that comes 
originally from string matching (e.g., the KMP
algorithm~\cite{Knuth1977} and the BM algorithm~\cite{Boyer1977}). The  optimization there is by pre-computing a skip value
table, and skipping
unnecessary matching trials accordingly. 

\section{Timed Pattern Matching}
We take an ``event-based'' formalization of timed pattern matching in~\cite{DBLP:conf/formats/WagaAH16,DBLP:conf/formats/WagaHS17}, unlike a ``state-based'' one in~\cite{Ulus2014,Ulus2016}. 
To represent a log of a real-time system, we employ a \emph{timed word},
which is a sequence of characters each of which is equipped with a real-valued timestamp.

\begin{mydefinition}
 [timed word]
 For an alphabet $\Sigma$, a \emph{timed word} over $\Sigma$ is a
 sequence
 $w = (a_1,\tau_1),(a_2,\tau_2),\dots,(a_n,\tau_n)\in
 (\Sigma\times\Rnn)^*$ satisfying
 $\tau_i \leq \tau_{i+1}$ for any $i \in [1,n-1]$.
\end{mydefinition}

We let $w|_{(t,t')}$ denote the restriction of $w$ to an interval $(t,t')$.
See~\cite{DBLP:conf/formats/WagaHS17} for details.

To represent a real-time specification, we employ a TA, which is an NFA equipped with timing constraints. 
Since a TRE~\cite{Asarin2002} can be translated to a
TA~\cite{Asarin2002}, we can also use a TRE as a
specification. The set of timed words accepted by a TA $\mathcal{A}$ is denoted by $L(\mathcal{A})$. 

Finally, our problem is formalized as follows.

\begin{mydefinition}
 [timed pattern matching]
 For a TA $\mathcal{A}$ and a timed word $w$, the \emph{timed
 pattern matching} problem asks for the set of \emph{matching intervals} 
 $\{(t,t')\mid w|_{(t,t')} \in L(\mathcal{A})\}$.
\end{mydefinition}


\section{\monaa---a MONitoring tool Accelerated by Automata}

We present a tool \emph{\monaa} for timed pattern matching.
In \monaa,  our \emph{timed FJS}
algorithm~\cite{DBLP:conf/formats/WagaHS17} is implemented.
It has the online property and enjoys the constant speedup by skipping,
 typically twice or three times faster than without skipping.
\monaa has two interfaces, the command-line interface \monaa and the C++ API
\libmonaa.

\paragraph*{Algorithm Description}
At the beginning of the timed FJS algorithm, it pre-constructs a \emph{skip value
table}. The table shows the number of matching trials to be skipped, depending on 
the observations obtained in the
 matching trial so far.  The original FJS algorithm is for string
matching~\cite{DBLP:journals/jda/FranekJS07}; there a skip value table is
constructed comparing strings, exploiting finiteness of the
pattern string. In
our timed FJS algorithm---where a pattern is an infinite set $L(\mathcal{A})$ of words rather than a single string---defining a \emph{finite} skip value table itself is a challenge. We use discrete states of TA for overapproximation, and construct a skip value table by checking emptiness of the intersection of the original TA $\mathcal{A}$ and its variant where the initial state is shifted. In this process we crucially rely on TA constructions such as \emph{zones}.


During the actual search, the timed FJS algorithm skips unnecessary matching
trials using the pre-constructed skip value table. We remark that the
runtime overhead of skipping is only by memory access and thus small.

\paragraph*{The Command-Line Interface}
In the command-line interface, \monaa reads a specification in either
:  a TA given in a file; or a TRE given as a command line argument. Reading a
timed word from the standard input, \monaa writes the result of the timed
pattern matching procedure to the standard output. Since \monaa reads the
timed word lazily, it can 
process a partial log provided by a system that is currently running. It can also notify a user of detection of matching behaviors before the whole matching is complete. 

\paragraph{The C++ API}:
We also provide a C++ API called \libmonaa. Because of the
\emph{modularity}, this API allows a user to write a program which
performs the timed pattern matching procedure as  part of the
program.
For example, one can implement a controller monitored in parallel,
and the monitor changes the control mode
when an unsafe behavior is detected.


In addition to the modularity, it also turns out that our C++ API is
beneficial for \emph{performance}. By hard-coding a TA in C++ code, 
we benefit from compiler optimization, and monitoring becomes faster.

\section{A Performance Comparison with Montre}

We compare the performance of \monaa with that of the existing tool Montre~\cite{DBLP:conf/cav/Ulus17}, by monitoring
 real-time behaviors of a Simulink model from an automotive domain.
The input timed words are generated from an automatic transmission
model~\cite{DBLP:conf/cpsweek/HoxhaAF14}.
The input specification is the following TRE or a corresponding TA
(modulo minor rewriting for readability).
\begin{align*}
\langle&(\text{g}_1\text{g}_2\text{g}_3\text{g}_4[\omega \ge 2500])
\lor(\text{g}_1\text{g}_2\text{g}_3[\omega \ge 2500]\text{g}_4)\\
\lor&(\text{g}_1\text{g}_2[\omega\ge 2500]\text{g}_3\text{g}_4)
\lor(\text{g}_1[\omega\ge 2500]\text{g}_2\text{g}_3\text{g}_4)\\
\lor&([\omega \ge 2500]\text{g}_1\text{g}_2\text{g}_3\text{g}_4)\rangle_{(0,
 10)} \\
 &\langle(\text{g}_3\lor\text{g}_4\lor[\omega< 2500]\lor[\omega\ge 2500])^+\rangle_{(1,1000)}
\end{align*}

This means that the gear changes from the first ($\text{g}_1$) to the
forth ($\text{g}_4$) and the engine rotation becomes high 
($[\omega\ge 2500]$) within 10 seconds, and in the next 1 second, the
gear keeps being the third $(\text{g}_3)$ or the forth $(\text{g}_4)$
but the velocity does not get high ($[v \geq 100]$).

We compared \monaa giving either a TRE or a TA, and a \libmonaa-based
timed pattern matching program (in which a TA is hard-coded), with
Montre's online and offline modes. Our
programs are compiled by GCC 7.1.0 with optimization flag -O3 and the
experiments are conducted on an Amazon EC2 c4.large instance (January
2018, 2 vCPUs and 3.75 GiB RAM) that runs Ubuntu 16.04.2 LTS (64 bit).

\begin{table}[tb]
 \newcommand{\logdate}{0129094951}
 \centering
 \caption{Execution time (sec.)}
 \vspace{-1em}
 \label{table:result_montre}
  \scalebox{0.65}{
   \pgfplotstabletypeset[
   sci,
   sci zerofill,
   multicolumn names, 
   columns={[index]0,[index]1,[index]2,[index]3,[index]4,[index]5},
   display columns/0/.style={
   column name=\begin{tabular}{c}Length of\\timed word\end{tabular}, 
   fixed,fixed zerofill,precision=0,
   },
   display columns/1/.style={
   column name=\begin{tabular}{c}\monaa\\(TRE)\end{tabular},
   precision=2},
   display columns/2/.style={
   column name=\begin{tabular}{c}\monaa\\(TA)\end{tabular},
   precision=2},
   display columns/3/.style={
   fixed,precision=2,
   column name=\begin{tabular}{c}\libmonaa\\(TA is \\ hard coded)\end{tabular}},
   display columns/4/.style={
   fixed,fixed zerofill,precision=2,
   column name=\begin{tabular}{c}Montre\\(online)\end{tabular}},
   display columns/5/.style={
   fixed,fixed zerofill,precision=2,
   column name=\begin{tabular}{c}Montre\\(offline)\end{tabular}},
   fixed,fixed zerofill,precision=2,
   every head row/.style={
   before row={\toprule}, 
   after row={\midrule} 
   },
   string replace={0}{},
   empty cells with={$< 0.01$},
   every last row/.style={after row=\bottomrule}, 
   ]{./figs/\logdate/execution_time.tsv}}
 \centering
 \caption{
Memory usage (kbytes)}
 \vspace{-1em}
 \label{table:result_montre_ram}
  \scalebox{0.60}{
   \pgfplotstabletypeset[
   sci,
   sci zerofill,
   multicolumn names, 
   columns={[index]0,[index]1,[index]2,[index]3,[index]5},
   display columns/0/.style={
   column name=\begin{tabular}{c}Length of\\timed word\end{tabular}, 
   fixed,fixed zerofill,precision=0,
   },
   display columns/1/.style={
   column name=\begin{tabular}{c}\monaa\\(TRE)\end{tabular},
   precision=0},
   display columns/2/.style={
   column name=\begin{tabular}{c}\monaa\\(TA)\end{tabular},
   precision=0},
   display columns/3/.style={
   fixed,precision=0,
   column name=\begin{tabular}{c}\libmonaa\\(TA is \\ hard coded)\end{tabular}},
   display columns/4/.style={
   fixed,fixed zerofill,precision=0,
   column name=\begin{tabular}{c}Montre\\(offline)\end{tabular}},
   fixed,fixed zerofill,precision=0,
   every head row/.style={
   before row={\toprule}, 
   after row={\midrule} 
   },
   every last row/.style={after row=\bottomrule}, 
   ]{./figs/\logdate/max_res_set.tsv}}
 \vspace{-1em}
\end{table}

The results of our experiments are in
Table~\ref{table:result_montre}--\ref{table:result_montre_ram}. 
Table~\ref{table:result_montre} shows that \libmonaa-based monitor
performs the fastest and the online mode of Montre performs the slowest.
We remark that \monaa constantly takes about 7 seconds extra when a
TRE is given. This is because of the translation
from a TRE to a TA, which does not affect the remaining procedure.
The execution time of \monaa grows only  linearly with respect to the length
of the input timed word, a characteristic desired for monitoring algorithms.

Table~\ref{table:result_montre_ram} shows that the memory usage of \monaa is
independent of the length of the timed word, while that of Montre
offline depends.

\section*{Acknowledgment}
\scriptsize
The authors are supported by
JSPS Grant-in-Aid 15KT0012. 
M.W.\ and I.H.\ are supported by JST ERATO HASUO Metamathematics for Systems Design Project (No.\
JPMJER1603),  and JSPS Grant-in-Aid No.\ 15K11984. 
K.S.\ is supported by JST PRESTO (No.\ JPMJPR15E5) and JSPS Grant-in-Aid No.\ 70633692.

\bibliographystyle{IEEEtran}

\end{document}